\documentclass[12pt]{article}
\usepackage{graphicx,amsmath}
\usepackage{units}
\usepackage{color}

\newlength{\dinwidth}
\newlength{\dinmargin}
\def\lapproxeq{\lower .7ex\hbox{$\;\stackrel{\textstyle                                                    
<}{\sim}\;$}}                                                    
\def\gapproxeq{\lower .7ex\hbox{$\;\stackrel{\textstyle                                                    
>}{\sim}\;$}}                                                    
\def\be{\begin{equation}}                                                    
\def\ee{\end{equation}}                                                    
\def\bea{\begin{eqnarray}}                                                    
\def\eea{\end{eqnarray}}

\def\GeV{\rm GeV}

\def\sh{\hat s}
\def\sh2{{\hat s}^2}

\def\CNLO{C^{\rm NLO}}

\def\MS{\overline{\rm MS}}
\def\Q{h}
\begin{document}

\begin{flushright}                                                    
IPPP/13/53  \\
DCPT/13/106 \\                                                    
\today \\                                                    
\end{flushright} 

\vspace*{0.5cm}

\begin{center}
{\Large \bf Treatment of heavy quarks in QCD}

\vspace*{1cm}
                                                   
E.G. de Oliveira$^{a,b}$, A.D. Martin$^a$, M.G. Ryskin$^{a,c}$ and A.G. Shuvaev$^{c}$  \\                                                    
                                                   
\vspace*{0.5cm}                                                    
$^a$ Institute for Particle Physics Phenomenology, University of Durham, Durham, DH1 3LE \\                                                   
$^b$ Instituto de F\'{\i}sica, Universidade de S\~{a}o Paulo, C.P.
66318,05315-970 S\~{a}o Paulo, Brazil \\
$^c$ Petersburg Nuclear Physics Institute, NRC Kurchatov Institute, Gatchina, St.~Petersburg, 188300, Russia \\          
                                                    
\vspace*{1cm}

\begin{abstract} 

We show that to correctly describe  the effects of the heavy-quark mass, $m_\Q$, in DGLAP evolution, it is necessary to work in the so-called `physical' scheme. In this way, we automatically obtain a smooth transition through the heavy-quark thresholds. Moreover, we show that to obtain NLO accuracy, it is sufficient to account
for the heavy-quark mass, $m_\Q$, just in the LO (one-loop) splitting function.  The use of the $\MS$ factorisation scheme is not appropriate, since at NLO  we deal with a mixture of quarks and gluon (that is, the mass of the heavy parton is not well-defined).
The formulae for the explicit $m_\Q$ dependence of the splitting functions, and for $\alpha_s$, are presented.

\end{abstract}                                                        
\vspace*{0.5cm}                                                    
                                                    
\end{center}

\section{Introduction  \label{sec:1}} 
The correct treatment of heavy quarks in an analysis of parton distributions is essential for precision measurements at hadron colliders. The up, down and strange quarks, with $m^2 \ll \Lambda^2_{\rm QCD}$, can be treated as massless partons. However, for charm, bottom or top quarks we must allow for the effects of their mass, $m_\Q$ with $\Q=c,b$ or $t$. The problem is that we require a consistent description of the evolution of parton distribution functions (PDFs) over regions which include {\it both} the  $Q^2 \sim m^2_{\Q}$ domain and the region $Q^2 \gg m^2_\Q$ where the heavy quark, $\Q$, can be treated as an additional massless quark. 

Let us briefly summarize how heavy quarks are treated in PDF analyses at present\footnote{A detailed review can be found in \cite{TT}.}. These analyses are performed in the $\MS$ scheme, in which the splitting and coefficient functions have been calculated  using dimensional regularisation. We will call this the `conventional' approach. Starting the evolution at a low scale we need consider only the three light quarks, taken as massless. As we evolve upwards we reach the charm quark threshold $Q^2=m_c^2$. We could choose to keep just the three light flavours as quark PDFs, and include all the effects of the charm quark and its mass $m_c$ in the coefficient functions. Historically, higher-order calculations of charm production were done in such a so-called Fixed-Flavour-Number-Scheme (FFNS)  \cite{FFNS}. 

 Unfortunately, a FFNS cannot be used far above the threshold. For $\ln Q^2 \gg \ln m^2_c$, the charm quark starts to participate in the evolution. Therefore in the FFNS coefficient functions, we have to sum up an infinite number of diagrams in order to reproduce inside these functions all the missing effects in the DGLAP evolution.
Indeed, higher-order contributions, $\alpha_s^n \ln^n(Q^2/m_c^2)$, do not decrease in comparison to lower-order terms, and perturbation theory breaks down.  Here, we call this a 3-flavour scheme (3FS). For higher $Q^2$ we should include the $c$-quark (which is taken as massless) in the evolution and so generate~\footnote{There are special processes where a 4-flavour set of partons is still necessary, see, for example \cite{MSTWFF}.} a 4FS giving reliable results for $m_c^2\lapproxeq Q^2 \lapproxeq m_b^2$, and 5FS giving reliable results for $m_b^2\lapproxeq Q^2 \lapproxeq m_t^2$ and so on.

Hence, we are led to a more general Variable-Flavour-Number-Scheme (VFNS), which is a composite of a sequence of $n_f$-flavour schemes, each with its own region of validity. As we pass through each transition {\it point}, $Q_{\rm trans}$ (usually taken as $Q^2=m_\Q^2$), the number of quarks active in the evolution increases from $n_f$ to $n_f+1$. So at a transition point we have two different sets of PDFs: the $n_f$-FS set for $Q^2 \le m_\Q^2$ and the $(n_f+1)$-FS set for $Q^2 \ge m_\Q^2$. The two sets have to be matched together in the transition region. The matching conditions are
\be
a_i^{n+1}(Q^2)~=~\sum_k A_{ik}(Q^2/m^2_\Q)~ \otimes~ a_k^n(Q^2),
\label{eq:1}
\ee
where $\otimes$ denotes the convolution
\be
A\otimes a~=~\int_x^1 \frac{dx'}{x'}~A(x')~a(x/x'),
\ee
where the PDF set ~ $a_i^n=g,q,\Q$,~ with 3 light quarks and $n-3$ heavy quarks $\Q$. We have suppressed the $x$ arguments in (\ref{eq:1}). The perturbative matrix elements $A_{ik}(Q^2/m_\Q^2)$ contain $\ln(Q^2/m_\Q^2)$ terms known to ${\cal O}(\alpha_s^2)$~\footnote{In general the matching may be performed at any $Q^2=c\cdot m^2_\Q$, where the value of $c$ is $c  \gapproxeq 1$. Recall however, that actually the matching (\ref{eq:1}) is done at  one, {\em fixed} point $Q^2=Q^2_{\rm trans}$, say $Q^2=m^2_\Q$.}. In summary, the various $n_f$ schemes are related to each other by perturbatively calculable transformation matrices between 
the PDFs and the coefficient functions\footnote{In analogous way the smooth behaviour of the coupling $\alpha_s(Q^2)$ is provided.}. 

Note that the matrix $A$ in (\ref{eq:1}) is not a square matrix; neglecting the NNLO correction there is some freedom in performing the matching at the transition points,
 which is exploited phenomenologically to ensure that the matching is as smooth as possible. The ACOT~\cite{ACOT} and RT~\cite{RT} prescriptions were early attempts to implement this matching. An important development was the use of the so-called the General Mass (GM)-VFNS, which allows an estimate of the suppression of the final-state phase space when  heavy flavour is produced. A rescaling variable
\be
\chi~=~x\left(1+\frac{(\sum M_f)^2}{Q^2}\right)
\label{eq:rescale}
\ee
is introduced,
where the sum is over the heavy particles produced in the final state. (For example, neutral-current heavy-flavour production has $\Q {\bar \Q}$ in the final state, whereas a charge-current process has a single $\Q$.) Then the convolution $C\otimes$PDF, with the corresponding coefficient function, $C$, should be integrated over the momentum fraction range $\chi < \xi <1$. Rescaling shifts the momentum fraction variable in the PDF, $a(\xi,\mu^2)$, to a higher value than in the zero-mass case.   This rescaling prescription \cite{ACOTchi1,ACOTchi2} is known as ACOT$\chi$.

The GM-VFNS is adopted in the MSTW \cite{MSTW}, CTEQ(CT10) \cite{CT10} and NNPDF \cite{NNPDF} global parton analyses, although each analysis uses its own variant. For example, MSTW use the formalism of \cite{thorne73}, while NNPDF use a prescription~\cite{FLNR} based on the fixed-order next-to-leading-log (FONLL) method.
For comparison the most recent FFNS analysis \cite{ABM} finds both the value of $\alpha_s(M_Z^2)$ and the size of the gluon PDF at large $x$, significantly smaller than those of the GM-VFNS analyses.

The VFNS is well justified at LO accuracy. Indeed, at LO, 
in each cell (loop) of the evolution diagram, the transverse momenta, $k_{ti}$, or the virtualities, $k^2_i$, are strongly ordered;  $k^2_i\gg  k^2_{i-1}$
and a large logarithmic integration $\int^{k^2_{i+1}}_{k^2_{i-1}}dk^{2}_i/k^{2}_i$ compensates the small value of the  QCD coupling $\alpha_s(k^2_i)$.  The contribution from a finite interval
of $\ln(k^2_i)$ (say, $k^2_i\sim k^2_{i+1}$) is considered as a NLO
correction to the Leading Log evolution since here (from this extra loop) we get a small $\alpha_s$ now unaccompanied by a large logarithm. In the same way, we have to treat the heavy-quark mass dependence, which comes only from the finite region of $\ln(k^2_i)$ (that is, from  $k^2_i\sim m^2_\Q$) {\bf as a NLO effect}. Correspondingly, the effect of the {\it running} mass is a NNLO contribution.

If we account for the NLO corrections within the VFNS, where at each threshold, $Q^2=m^2_\Q$  we just increase the number of light active quarks by 1
(but each type of quark is considered as massless in the evolution), then, as mentioned above, we get JUMPs in the splitting functions (and kinks in $\alpha_s$) when the value of $n_f$ is changed. This behaviour is compensated by the matching condition (\ref{eq:1}).
 The effect of the kink is calculated and added to the NLO PDFs in 
 such a way as to provide the correct behaviour for $Q^2 \gg m^2_\Q$, assuming that there is only one threshold in this interval of evolution. The remaining kink in the derivative may be considered as a NNLO effect (and, in its turn, it can be compensated
for in the region $Q^2 \gg m^2_\Q$ at the NNLO level,  again  assuming that there is only one threshold in this interval of evolution).

The GM-VFNS  allows us to correctly reproduce the evolution in a large
$\ln Q^2$ interval, but it cannot describe precisely the behaviour in the regions around the heavy-quark thresholds $Q^2\sim m^2_\Q$.
Such an approach does not remove the jumps in the splitting functions at the transition point, $Q=Q_{\rm trans}$. The kinks in these domains are only compensated in some {\em average} sense.

In this paper, we propose a completely different, physically-motivated, approach, which automatically results in a {\it smooth} behaviour of the PDFs, the coefficient functions and of $\alpha_s$ as the scale $\mu^2$ passes through each heavy-quark threshold. In this `physical' approach the partons which occur in the Feynman diagrams are the basic entities. However, before we describe our approach, a comment about the conventional $\MS$ scheme is needed.

It was shown in \cite{OMR1} that the NLO coefficient functions, $\CNLO$,  
obtained within the `conventional' $\MS$ prescription using the dimensional ($D=4+2\epsilon$) regularization, are different from the results calculated in the `physical' approach of working in normal $D=4$ space where the infrared divergency is removed  by an appropriate subtraction of the contribution, $C^{\rm LO}\otimes P^{\rm LO}$, generated by the iteration of LO evolution. 

The above difference, $\Delta C$, is due to an $\epsilon/\epsilon$ contribution coming from very large (non-physical) distances. 
It can be written as the convolution 
\be
\Delta C_{ik}=C^{\rm LO}_{ik'}\otimes \delta_{k'k}(z),
\label{eq:cf}
\ee
 where $\delta_{ik}(z)$ denotes the part of the LO splitting functions that is proportional to $\epsilon$
\be
P_{ik}(z)=P_{ik}^{\rm LO}(z)+\epsilon\delta_{ik}(z).
\ee
This contribution can be absorbed in the redefinition of the partons, $a(x,\mu^2)=xg,\ xq,\ x\Q$ 
\be
a^{\rm phys}(x,\mu^2)=a^{\overline{\rm MS}}(x,\mu^2)+\frac{\alpha_s}{2\pi}\int
dz\sum_b\delta_{ab}(z)~
b^{\overline{\rm MS}}(x/z,\mu^2)\ .
\label{eq:newparton}
\ee
Correspondingly, there is a difference between the NLO splitting functions in DGLAP evolution equation obtained in the conventional $\MS$ scheme and the physical approach, see~\cite{OMR2} .
 As seen from (\ref{eq:newparton}), at NLO, the conventional 
 $\overline{\rm MS}$ partons are `rotated' with respect to the physical 
 partons by some angle. In particular, the singlet-quark distribution
gets an admixture of gluons\footnote{This `rotation' is compensated
by a corresponding rotation back arising from the difference in the splitting functions. For heavy quarks we do not have an infrared problem. However, in general, using the conventional $\MS$ scheme, we do not know the mass of the parton that we are dealing with. It is therefore difficult to account for mass effects in the $\MS$ scheme.}.
 This mixture greatly complicates the calculations of heavy-quark 
 mass effects, and any other Feynman graph calculations, beyond  LO, in the $\MS$ scheme.
By working in the `physical' approach, where we calculate the explicit  $m_\Q$ dependence of the DGLAP splitting functions (and of $\alpha_s$), we obtain a well-defined, and simplified treatment of heavy-quark mass effects to NLO accuracy.

In fact, in the present paper we show how to account for the heavy-quark mass {\it already} in the LO splitting function. To do this we calculate the explicit $m^2_\Q/Q^2$ dependence of the derivatives of the PDFs, $\partial a(x,Q^2)/\partial\ln Q^2$; instead of using the conventional splitting functions, which only depend on  $Q^2$  via the running coupling $\alpha_s(Q^2)$. In Section \ref{sec:2} the corresponding splitting functions
are calculated from the one-loop (i.e. LO) Feynman diagrams.  We  discuss 
whether, accounting for the mass effect already at the LO level, we have to correct the usual NLO splitting and the coefficient functions. In Section \ref{sec:3} we obtain
 an analogous LO formula which gives
 the effects of the heavy-quark masses to the running of $\alpha_s$ at NLO. This provides a smooth behaviour of $\alpha_s$ across the heavy-quark thresholds.
All the calculations are done in the `physical' scheme; so in the Appendix we present the formulae to provide the `rotation' from $\overline{\rm MS}$
 to physical scheme, and vice versa.
    We present our conclusions in Section \ref{sec:4}.

\section{Heavy-quark mass effects already included at LO   \label{sec:2}}
\begin{figure}
\begin{center}
\vspace*{-3.cm}
\includegraphics[height=8cm]{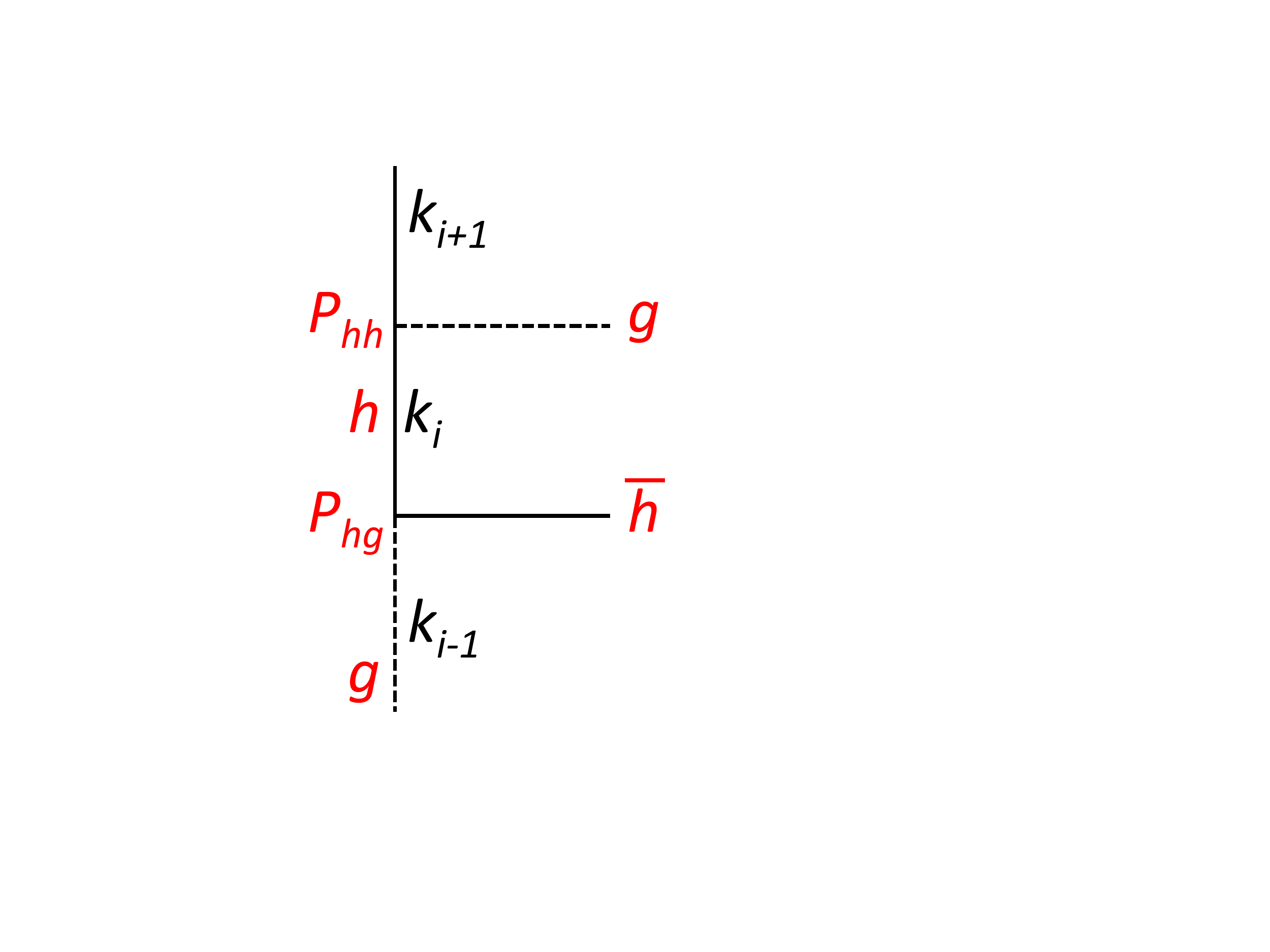}
\vspace*{-2.2cm}
\caption{\sf Part of the parton evolution chain which contains the $g \to \Q\bar{\Q}$ transition} 
\label{fig:1}
\end{center}
\end{figure}
Since the heavy-quark mass effects come only from a finite interval of the $\ln Q^2$ evolution, to reach the NLO accuracy it is sufficient to account for $m_\Q$ only in the LO diagrams. We will see that keeping the mass in the NLO (two-loop) graphs leads to a NNLO correction. As usual we use the axial gauge, where only the ladder (real emission) and the self-energy (virtual-loop contribution) diagrams  give Leading Logarithms. Actually, for  real emission  we need to consider only the `gluon-to-heavy quark'
splitting function.  Indeed the heavy-quark mass effects can be identified in the following subset of integrations
\be
...\int\frac{dk^2_{i-1}}{k^2_{i-1}}\int\frac{dk^2_i~ k_i^2}{(k^2_i+m_\Q^2)^2}\int\frac{dk^2_{i+1}}{k^2_{i+1}} ...
\label{eq:5}
\ee
corresponding to the part of the parton chain containing the $g \to \Q \bar{\Q}$ transition, as shown in Fig. \ref{fig:1}. The $k^2$'s are the virtualities of the $t$-channel partons, and the heavy-quark mass effects enter in the $k_i^2$ integration that results from the $g \to \Q\bar{\Q}$ transition. The kinematics responsible for the LO result are when the virtualities are strongly ordered (...$k^2_{i-1}\ll k^2_i \ll k^2_{i+1}$...). If two of the partons have comparable virtuality, $k^2_j \sim k^2_{j+1}$, then we lose a $\ln Q^2$ and obtain a NLO contribution of the form $\alpha_s(\alpha_s\ln Q^2)^{n-1}$ for $n$ emitted partons.

At first sight it appears that $m^2_\Q$ should also have been retained in the integration over the heavy-quark line with virtuality $k_{i+1}$. However, the heavy quark was produced at $Q^2 \sim m^2_\Q$ via the $g \to \Q$ splitting. Due to the strong ordering $k^2_{i+1} \gg k_i^2$ in the evolution chain, we have $k^2_{i+1} \gg m^2_\Q$, and so we may neglect $m^2_\Q$ in the $k_{i+1}^2$ integration; otherwise this would be the NNLO effect. 

Note that in our NLO calculations, described below, we use a fixed number $m_\Q(m_\Q)$ for the heavy quark mass\footnote{Strictly speaking we may choose any {\it reasonable} fixed value for $m_\Q$, say $m_c(1.4$ GeV), so that the NNLO correction is not large,}. All the effects of the running quark mass should be regarded as part of the NNLO corrections.

\subsection{Quark mass effects in the LO splitting functions}
We are now in a position to calculate the heavy-quark mass effects in real LO $\Q\ ({\bar \Q})$ production which  determines the explicit $m^2/\Q^2$ dependence
of $P_{\Q g}$. This, in turn, allows us to account for the $m_\Q$ dependence in the heavy-quark virtual-loop contribution (that is in the gluon self-energy), which gives an additional term in $P_{gg}$, which is proportional to $ \delta(1-z)$. Recall that the full $g\to g$ splitting function has the form
\be
\label{eq:delta-g}
P_{gg}(z)=P^{\rm real}_{gg}(z)-\delta(1-z)\int_0^1 \left(z'P^{\rm real}(z')+\sum_f P_{qg}(z')\right)dz'\ ,
\ee
where the $\sum_f$ includes the summation over different type of quarks.
Correspondingly
\be
\label{eq:delta-q}
P_{\Q \Q}(z)~=~P_{\Q \Q}^{\rm real}(z)-\delta(1-z)\int_0^1 P^{\rm real}_{g\Q}(z')dz'\ .
\ee

To determine the $m^2_{\Q}/Q^2$ dependence of $P_{\Q g}$ we must calculate  the one-loop ladder (heavy-quark box\footnote{Better to say ``heavy-quark triangle'', since the upper line with the largest $k_t$ at LO is treated as a `point-like operator'. }) diagram. We denote the virtuality of the $t$-channel heavy quark $\Q$ by $k_i^2$, as in Fig. \ref{fig:1}. Now strong-ordering means that the
virtuality of incoming gluon $k^2_{i-1} \ll k^2_i$ and that $k^2_i\ll k^2_{i+1}$. We find that the $m_\Q$ dependence of the LO `gluon to heavy-quark' splitting function, $P_{\Q g}$, is
\be
\label{eq:g-to-Q}
P_{\Q g}(z,Q^2)=T_R\left([z^2+(1-z)^2]\frac{Q^2}{m^2_\Q+Q^2}~+~\frac{2m^2_\Q Q^2 z(1-z)}{(Q^2+m^2_\Q)^2}\right)\Theta \left(Q^2-\frac{zm^2_\Q}{1-z}\right)\ .
\ee
where $T_R=1/2$. The first term is the usual LO splitting function $P_{\Q g}$ modified by a factor, $Q^2/(Q^2+m^2_\Q)$, which tends to 1 for  $Q^2\gg m^2_\Q$, while for low $Q^2$, $Q^2\ll m^2_\Q$, this contribution becomes negligible. The second term, proportional to $m^2_\Q$, accounts for the possibility to flip the helicity in the heavy-quark loop. It dies out for $Q^2\gg m^2_\Q$. Finally, the $\Theta$ function accounts for the correct kinematics of heavy-quark production. We need energy to put the heavy-quark on-mass-shell. This leads to a minimum value of the (longitudinal part of) $Q^2$.

Simultaneously we have to include heavy-quark loops in the gluon self-energy, as was mentioned in (\ref{eq:delta-g}). That is, we must add a term to the gluon-gluon splitting function, $P_{gg}$,
\be
\delta P_{gg}=-\delta(1-z)\sum_\Q \int_0^{z_\Q}P_{\Q g}(z',Q^2)dz'\ ,
\ee
 where the upper limits of integration, $z_\Q =Q^2/(Q^2+m^2_\Q)$, are determined by the
 $\Theta$ function in (\ref{eq:g-to-Q}).

For completeness, and to provide the smooth behaviour in all the LO splitting functions, we present the other two LO kernels which involve the heavy quark.
An analogous calculation for the $\Q\to \Q$ splitting gives
\be
P^{\rm real}_{\Q\Q}(z,Q^2)=C_F\left(\frac{1+z^2}{1-z}\frac{Q^2}{m^2_\Q+Q^2}~+~\frac{z(1-3z)}{1-z}\frac{Q^2m^2_\Q}{(Q^2+m^2_\Q)^2}\right)
\ee
and for the $\Q \to g$ transition
\be
P_{g\Q}(z,Q^2)=C_F\left(\frac{1+(1-z)^2}z\frac{Q^2}{m^2_\Q+Q^2}~+~\frac
{z^2+z-2}z\frac{Q^2m^2_\Q}{(Q^2+m^2_\Q)^2}\right)\Theta\left(Q^2-\frac{zm^2_\Q}{1-z}\right)\ .
\ee
 
 We may summarize the LO evolution
equations in the symbolic form
\bea
\label{eq:b4}
\dot{g} & = & P_{gg} \otimes g \: + \: \sum_q~P_{gq} \otimes q \:
+
\: \sum_\Q P_{g\Q} \otimes \Q \nonumber \\
\dot{q} & = & P_{qg} \otimes g \: + \: P_{qq} \otimes q \\
\dot{{\Q}} & = & P_{
\Q g} \otimes g \: + \: P_{\Q\Q} \otimes \Q \nonumber
\eea
where $q = u,d,s$ denotes the light quark density functions and
$\Q=c,b,t$ are the heavy-quark densities.  We have abbreviated $P^{\rm LO}$ by $P$, and
$\dot{a} = (2 \pi/\alpha_S) \partial a/\partial \ln Q^2$.

Note that there are evolution equations, (\ref{eq:b4}), for {\em all}~  type of partons (including  heavy quarks) just starting from $Q_0$. The input heavy-quark distribution $\Q(x,Q^2_0)$ should be treated as an `intrinsic' PDF introduced in~\cite{Brod}. Of course, at low $Q^2 \ll m^2_\Q$ the corresponding splitting functions are strongly suppressed by the small value of the ratio $Q^2/m^2_\Q$. So, actually the evolution of the heavy quark will start somewhere in the region $Q^2\simeq m^2_\Q$.

\subsection{Quark mass effects in NLO diagrams}
It turns out that to include heavy-quark mass effects in NLO evolution we do not
need to modify the usual NLO splitting functions. 
In the absence of intrinsic heavy quark, we only have to take
$m_\Q$ into account in $P_{\Q g}$ and then only in the LO part
$P_{\Q g}^{(0)}$.  (Of course, as a consequence, we must adjust the
virtual corrections to $P_{gg}$).  The argument is as follows.

The $k_i^2$ integral of (\ref{eq:5}) written with NLO accuracy, has the form
\be
\int \: \frac{dk_{i}^2 \: A(k_{i}^2,k^2_{i+1},m^2_\Q ,z)}{(k_{i}^2 + m_\Q ^2)^2} \; =
\; \int \: A_1(z)\frac{d (k_{i}^2 + m_\Q^2)}{(k_{i}^2 + m_\Q^2)} \: +
\: \int \: A_2(z)\frac{m_\Q^2 \: dk_{i}^2}{(k_{i}^2 + m_\Q^2)^2}+
\: \int \: A_3(z)\frac{dk_{i}^2}{k_{i+1}^2}.
\label{eq:b6}
\ee
The first term gives the leading logarithm contribution.  To be specific we have 
\be
\int_{k_{i - 1}^2}^{Q^2} \: \frac{d k^2}{(k^2 + m_\Q^2)} \; =
 \; \ln \: \frac{Q^2+m^2_\Q}{m_\Q^2}
\label{eq:b7}
\ee
for $k_{i - 1}^2 \ll m_\Q^2$. 
Both the second term in
(\ref{eq:b6}), which is concentrated in the region $k_{i}^2 \sim
m_\Q^2$, and the third term, which is concentrated near the upper limit, at $k^2_i\sim k^2_{i+1}$, give  non-logarithmic contributions.

In the axial gauge the two first terms on the right-hand-side of (\ref{eq:b6}) come only from the pure ladder (and the corresponding self-energy) diagrams, from the region of $k^2_i \ll k^2_{i+1}$. That is, these two terms are exactly the same as those generated by  LO$\otimes $LO evolution, in which we have already accounted for the $m_\Q$ effects. To avoid double counting, we have to subtract these contributions from (\ref{eq:b6}). Thus the true NLO contribution is given by the third term only, in which we can omit the $m_\Q$ dependence since: (a) 
$k^2_{i+1}\gg m^2_\Q$, and, (b) these order of ${\cal O}(m^2_\Q /k^2_{i+1})$ terms kill the large logarithm in the further $\int dk^2_{i+1}/k^2_{i+1}$ integration.
 That is, at NLO accuracy we can use the old, well-known, NLO splitting 
 functions $P^{(1)}_{ik}(z)$. If we were to account for the mass effect in  
 $P^{(1)}_{ik}(z)$, then we would be calculating a NNLO 
 correction\footnote{Before proceeding to NNLO, a phenomenological way to provide very smooth behaviour of the 
 NLO contribution would be to multiply the `heavy-quark' NLO terms (that is, those NLO terms 
 which are proportional to $n_\Q$)  simply by the factor $Q^2/(Q^2+m^2_\Q)$.}.

In summary, there are no heavy-quark mass effects in the NLO splitting functions. Only LO $P^{(0)}_{\Q g}$ needs to be modified in order to reach the NLO accuracy in the absence of intrinsic heavy quark.

\subsection{NLO coefficient functions}
Recall that the NLO coefficient function, $C^{\rm NLO}(z,Q^2)$, is calculated assuming that the incoming parton virtuality, $k^2_n$ is much less than the scale $Q^2$, so that the integral $\int^{Q^2} dk^2_n/k^2_n$ has a logarithmic form. This means that at NLO accuracy we may neglect the virtuality of the incoming parton. Moreover, in our `physical' scheme there is no mixture of different types of partons (like those generated by
(\ref{eq:1}) in conventional VFNS). As a consequence there
is no change to the NLO coefficient functions.

Note also that in the physical scheme we consistently use the $x$ variable as the light cone momentum fraction and it not necessary to introduce rescaling described in (\ref{eq:rescale}), and the subsequent text.. That is, in the physical scheme we deal with quantities which have a clear physical interpretation.

\section{Smooth $\alpha_s$ evolution across a heavy-quark threshold  \label{sec:3}}

\begin{figure}
\begin{center}
\includegraphics[height=8cm]{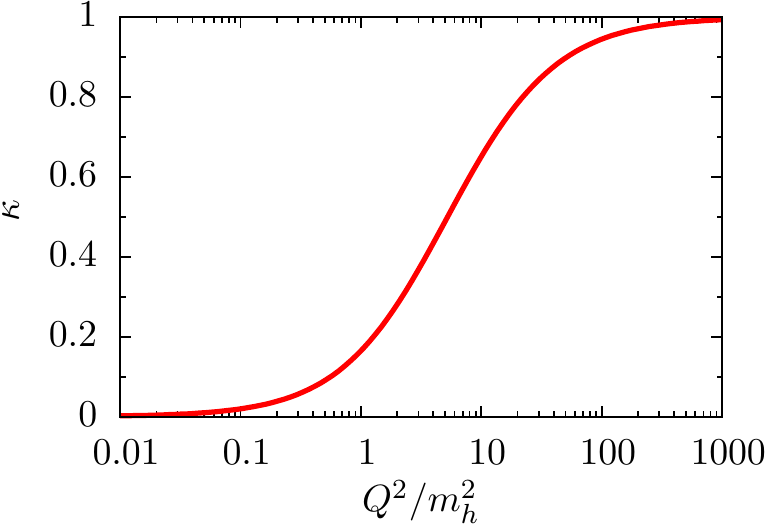}
\vspace*{-0.5cm}
\caption{\sf The contribution of a heavy quark to the running of $\alpha_s$, showing a smooth behaviour across the heavy-quark threshold. If $\kappa=1$, the heavy quark acts as if it were massless.} 
\label{fig:2}
\end{center}
\end{figure}
In analogous way we account for the heavy-quark mass effect in the QCD coupling $\alpha_s(Q^2)$. The running of the coupling to NLO is given by
\be
\frac{d}{d\ln Q^2}\left(\frac{\alpha_s}{4\pi}\right)~=~-\beta_0\left(\frac{\alpha_s}{4\pi}\right)^2  -\beta_1\left(\frac{\alpha_s}{4\pi}\right)^3 ,
\ee
where the $\beta$-function coefficients are
\be
\beta_0(n_f)=11-\frac{2}{3} n_f,~~~~~~~~   \beta_1(n_f)=102-\frac{38}{3} n_f. 
\ee
To determine the effect of a heavy quark mass in the running of $\alpha_s$ at NLO, it is sufficient to calculate
the `gluon to heavy-quark' loop insertion (that is, the gluon self energy) to gluon propagator. This fermion loop insertion is responsible for the $-(2/3)n_f$
term in the LO $\beta$-function.
In this way we find that, instead of  changing $n_f$ from 3 to 4 (at $Q^2=m^2_c$), and from 4 to 5 (at $Q^2=m^2_b$), we must include in $n_f$ a term 
\be
\label{alpha}
\kappa(r)~=~\left[1-6r+12\frac{r^2}{\sqrt{1+4r}}\ln\frac{\sqrt{1+4r}+1}{\sqrt{1+4r}-1}\right]\ ,
\ee 
for each heavy quark, where $r\equiv m^2_\Q /Q^2$. In Fig. \ref{fig:2}  we plot $\kappa$ as a function of $Q^2/m^2_{\Q}$.  As expected, $\kappa \to 1$ at large $Q$, where the heavy quark acts as if it were massless, but even for  $Q^2 \simeq 10 m^2_\Q$ we see that the effects of $m_\Q$ are very important. For $Q^2\ll m^2_\Q$ it vanishes as $1/5r=Q^2/5m^2_\Q$, so there is only a small heavy-quark contribution to $n_f$ for $Q<m_\Q$.

\begin{figure}
\begin{center}
\includegraphics[height=8cm]{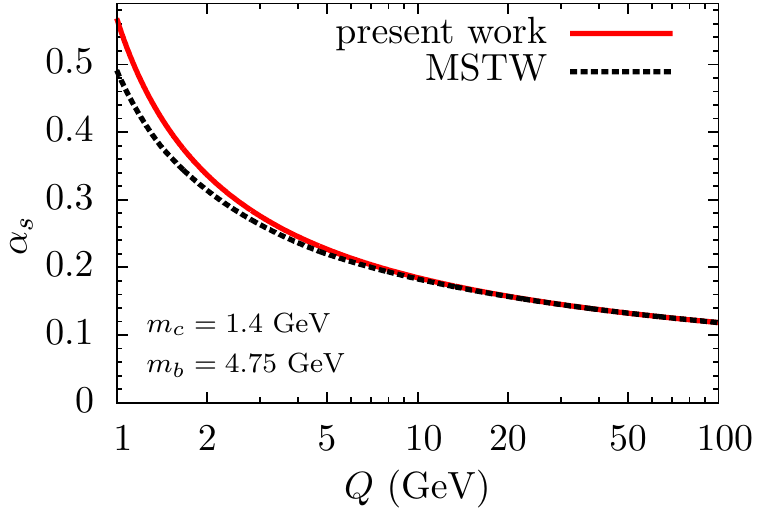}
\includegraphics[height=8cm]{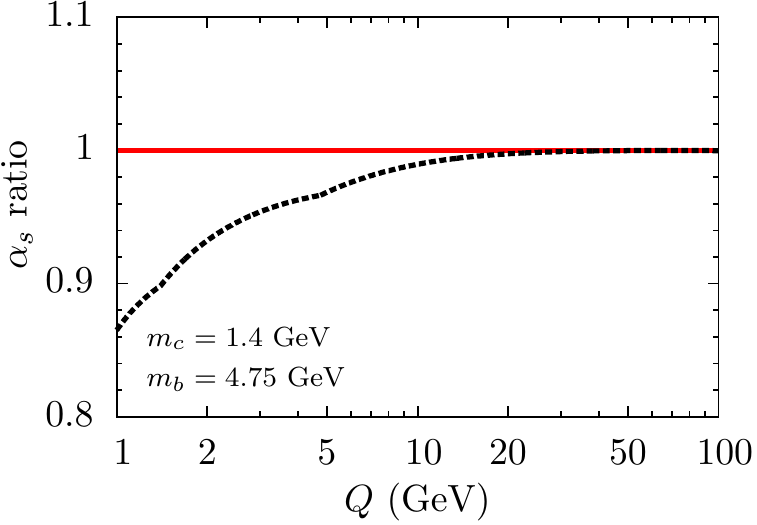}
\vspace*{-0.5cm}
\caption{\sf (a) The running of $\alpha_s$ at NLO: the continuous curve is obtained with the effects of the heavy-quark masses $m_c,~m_b$ included, and the dashed curve is that used by MSTW. Both evolutions are normalised to $\alpha_s(M_Z^2)=0.12$. (b) The ratio of the above two evolutions of $\alpha_s$.} 
\label{fig:3}
\end{center}
\end{figure}
In Fig. \ref{fig:3} we compare the evolution of $\alpha_s$ in which the effects of the heavy-quark masses are included, with an evolution assuming all quarks are massless. In the latter case a prescription has been used to ensure that $\alpha_s$ is continuous across the heavy-quark thresholds. Different prescriptions are possible, but it is not possible to make the derivative also continuous, as can be seen from Fig. \ref{fig:3}(b).  Indeed, with massless evolution, different reasonable prescriptions can lead to a difference of more than 0.5$\%$ in going from $Q^2 \sim 20~\GeV^2$ up to $Q^2=M_Z^2$, see the Appendix in \cite{MSTWFF}. However, when the heavy quark masses are properly accounted for, we see that the difference over this interval is about $4\%$, and in fact up to  14$\%$ starting from $Q^2=1$ GeV$^2$. The fact that the $\alpha_s$ curve, obtained with mass effects included, lies consistently above that for massless evolution in Fig. \ref{fig:3}(a) follows from the behaviour of $\kappa$ in Fig. \ref{fig:2} and that we have required both curves to have $\alpha_s(M_Z^2)=0.12$.

 \section{Conclusions \label{sec:4}}
In order to account for the effects of the heavy-quark mass, $m_\Q$, in DGLAP evolution, and to provide a smooth transition through the heavy-quark threshold regions, we include the $m_\Q$ dependence already in the LO (one-loop) splitting functions. We show that this modification of the LO splitting functions already provides NLO accuracy; there is no need to modify the known NLO splitting and coefficient functions. The crucial difference of our approach with those of the conventional FFNS or VFNS, is the fact that the heavy-quark mass is included directly in the splitting functions; that is, the heavy-quark mass is retained throughout the evolution. The presence of the quark mass in the splitting function automatically suppresses the evolution of the heavy quark at low scales, $Q^2 \ll m^2_\Q$, while at  large $Q^2 \gg m^2_\Q$ the massless limit is restored.

 To express it another way, by explicitly calculating the appropriate Feynman diagrams, keeping the heavy-quark mass dependence, we obtained the corresponding expressions for the LO splitting functions and the running QCD coupling $\alpha_s$. In this way, we have determined the full $m^2_\Q /Q^2$ behaviour of DGLAP evolution at NLO.

The idea to account for the heavy-quark mass already in the splitting function was proposed in~\cite{MRRS}.\footnote{ A similar splitting function which depends on quark mass was presented in earlier work by \cite{GHR}. An analogous result for QED may be found in \cite{BFK}.} However, there, the conventional $\MS$ factorization scheme was used and the splitting function still has some irregularity, since the heavy-quark part was included only at large enough $Q^2$ above the heavy-quark threshold. It was shown in~\cite{OS} that, in this form, the resulting physical cross sections are not different from those obtained in the conventional VFNS approach. On the other hand, in the present paper, we work in the `physical' scheme including the effect of the heavy-quark mass consistently starting from the input scale $Q_0$ of the DGLAP evolution. Therefore, there are no irregularities at the heavy-quark thresholds. All the formulae at NLO level are quite simple.  The generalization to NNLO is straightforward -- we need to account for $m_\Q$ in the two-loop (NLO) diagrams.

We emphasize the advantage of using the `physical' scheme where we deal with the true physical quantities: there is no mixture of the partons of different types, and no `rescaling' of the  $x$ variable as in (\ref{eq:rescale}). Our $x$ is just the light-cone momentum fraction. Thus the mass of each parton is well-defined. In contrast in the $\MS$ factorization scheme, at  NLO level, we deal with some mixture of partons - for example the singlet quark distribution has an admixture of gluons, and so on.  Recall that also in NLO Monte Carlos, where the quantum numbers of each parton must be correctly defined, an alternative scheme to the $\MS$ scheme is used~\cite{MC}.  Our approach should also be useful to
compare Monte Carlo event generators, in which parton
radiation is similarly performed including heavy quark masses, with the
analytical results.

To summarize, the treatment of heavy-quark mass effects is perturbatively calculable in QCD with no ambiguity\footnote{There is the possibility of a small ${\cal O}(1/m_\Q^2)$ non-perturbative `intrinsic' heavy-quark component in the starting heavy-quark distributions to the DGLAP evolution.}, with the heavy-quark masses as free parameters. It is not necessary to adopt one of the GM-VFNSchemes (or a FFNS). In the `physical' scheme that we introduce, there is a smooth behaviour of all quantities across the heavy-quark thresholds.  Clearly, Fig. \ref{fig:3}(b), for example, shows that a new global analysis of data is essential to determine the PDFs of the proton. However, first, we must complete the calculation of all the splitting and coefficient functions in the physical scheme; this is underway.

\section*{Appendix}
At NLO accuracy the relation between the $\overline{\rm MS}$ and the physical parton distributions are given by (\ref{eq:newparton}), where the long-distances part of the NLO coefficient function $\Delta C$ originates from the term  proportional to $\epsilon$  in the LO splitting
functions 
\be
P_{ab}(z)=P_{ab}^{\rm LO}(z)+\epsilon\delta_{ab}(z). 
\label{eq:99}
\ee
The $\epsilon$-
 dependent term, $\delta_{ab}$ is known (see for example~\cite{CG}). However, in comparison with the results listed in \cite{CG}, we have to add a contribution of pure kinematical origin.
Indeed, in $D=4+2\epsilon$ space the logarithmic integration $\int dk^2_t/k^2_t$ is replaced by $\int d^{2+2\epsilon}k_t/k^2_t\propto (1/\epsilon)(k^2_t)^\epsilon$. If expressed in terms of the virtuality variable, this phase-space factor $(k^2_t)^\epsilon$ reads
\be
(k^2_t)^\epsilon=(k^2(1-z))^\epsilon=1+\epsilon\ln k^2+\epsilon\ln(1-z)\ .
\ee
The last term in this expansion leads to an additional contribution to $\delta_{ab}(z)$ of (\ref{eq:99}) of the form $P^{\rm LO}_{ab}(z)\ln(1-z)$.
Thus we obtain
\be
P^{\rm real}_{qq}(z)=C_F\left[\frac{1+z^2}{1-z}(1+\epsilon\ln(1-z))
+\epsilon(1-z)
\right]\ , 
\ee
\be
P_{qg}(z)=T_R\left[(z^2+(1-z)^2)(1+\epsilon\ln(1-z)) +\epsilon 2z(1-z)\right]\ ,
\ee
\be
P_{gq}(z)=C_F\left[\frac{1+(1-z)^2}z(1+\epsilon\ln(1-z))+\epsilon z\right]\ ,
\ee
\be
P^{\rm real}_{gg}(z)=2C_A\left[\left(\frac z{1-z}+\frac{1-z}z+z(1-z)\right)(1+\epsilon\ln(1-z))
\right]\ .
\ee
To be complete, recall also the relation between the $\MS$ NLO coefficient functions and those in the physical scheme. As was mentioned already in Section \ref{sec:1},
\be
C^{\rm NLO}_a({\rm phys})=C^{\rm NLO}_a(\MS)-\sum_iC^{\rm LO}_i\otimes\delta_{ia},
\ee
see (\ref{eq:cf}).

\section*{Acknowledgements}

We thank Robert Thorne for valuable discussions. EGdO and MGR thank the IPPP at the University of Durham for hospitality. This work was supported by the grant RFBR 11-02-00120-a
and by the Federal Program of the Russian State RSGSS-4801.2012.2;
and by FAPESP (Brazil) under contract 2012/05469-4.

\end{document}